\documentclass[prx,twocolumn,english,superscriptaddress,floatfix,longbibliography]{revtex4-2}

\usepackage{ulem}
\usepackage[dvipsnames]{xcolor}
\usepackage{graphicx}
\usepackage{dcolumn}
\usepackage{bm}
\usepackage{amssymb}
\usepackage{amsmath}
\usepackage{physics}
\usepackage{float}

\begin{document}
\preprint{APS/123-QED}


\title{Anomalous metallic phase and reduced critical current in superconducting nanowires due to inverse proximity effect}

\affiliation{Departamento de F\'{i}sica de la Materia Condensada, Universidad Aut\'{o}noma de Madrid, Madrid, Spain}
\affiliation{Departamento de F\'{i}sica Te\'{o}rica de la Materia Condensada, Universidad Aut\'{o}noma de Madrid, Madrid, Spain}
\affiliation{Condensed Matter Physics Center (IFIMAC), Universidad Aut\'{o}noma de Madrid, Madrid, Spain}
\affiliation{Instituto de Ciencia de Materiales de Madrid (ICMM), Consejo Superior de Investigaciones Cientificas (CSIC),
Sor Juana Ines de la Cruz 3, 28049 Madrid, Spain}
\affiliation{Center for Quantum Devices, Niels Bohr Institute, University of Copenhagen, Madrid, Spain}
\affiliation{Laboratorio de Transporte Cu\'{a}ntico, Unidad Asociada
UAM/ICMM-CSIC, Madrid, Spain}

\author{G. M. Oliveira}
\thanks{These authors have contributed equally to this work.}
\affiliation{Departamento de F\'{i}sica de la Materia Condensada, Universidad Aut\'{o}noma de Madrid, Madrid, Spain}
\affiliation{Condensed Matter Physics Center (IFIMAC), Universidad Aut\'{o}noma de Madrid, Madrid, Spain}

\author{G. O. Steffensen}
\thanks{These authors have contributed equally to this work.}
\affiliation{Instituto de Ciencia de Materiales de Madrid (ICMM), Consejo Superior de Investigaciones Cientificas (CSIC),
Sor Juana Ines de la Cruz 3, 28049 Madrid, Spain}

\author{I. Casal Iglesias}
\affiliation{Departamento de F\'{i}sica de la Materia Condensada, Universidad Aut\'{o}noma de Madrid, Madrid, Spain}
\affiliation{Condensed Matter Physics Center (IFIMAC), Universidad Aut\'{o}noma de Madrid, Madrid, Spain}

\author{M. G\'{o}mez}
\affiliation{Departamento de F\'{i}sica de la Materia Condensada, Universidad Aut\'{o}noma de Madrid, Madrid, Spain}
\affiliation{Condensed Matter Physics Center (IFIMAC), Universidad Aut\'{o}noma de Madrid, Madrid, Spain}

\author{A. Ibabe}
\affiliation{Departamento de F\'{i}sica de la Materia Condensada, Universidad Aut\'{o}noma de Madrid, Madrid, Spain}
\affiliation{Condensed Matter Physics Center (IFIMAC), Universidad Aut\'{o}noma de Madrid, Madrid, Spain}

\author{T.Kanne}
\affiliation{Center for Quantum Devices, Niels Bohr Institute, University of Copenhagen, Copenhagen, Denmark.}

\author{J. Nyg\aa rd}
\affiliation{Center for Quantum Devices, Niels Bohr Institute, University of Copenhagen, Copenhagen, Denmark.}

\author{R. Aguado}
\affiliation{Instituto de Ciencia de Materiales de Madrid (ICMM), Consejo Superior de Investigaciones Cientificas (CSIC),
Sor Juana Ines de la Cruz 3, 28049 Madrid, Spain}
\affiliation{Laboratorio de Transporte Cu\'{a}ntico, Unidad Asociada
UAM/ICMM-CSIC, Madrid, Spain}
\author{A. Levy Yeyati}
\affiliation{Departamento de F\'{i}sica Te\'{o}rica de la Materia Condensada, Universidad Aut\'{o}noma de Madrid, Madrid, Spain}
\affiliation{Condensed Matter Physics Center (IFIMAC), Universidad Aut\'{o}noma de Madrid, Madrid, Spain}
\affiliation{Laboratorio de Transporte Cu\'{a}ntico, Unidad Asociada
UAM/ICMM-CSIC, Madrid, Spain}

\author{E. J. H. Lee}
\email{eduardo.lee@uam.es}
\affiliation{Departamento de F\'{i}sica de la Materia Condensada, Universidad Aut\'{o}noma de Madrid, Madrid, Spain}
\affiliation{Condensed Matter Physics Center (IFIMAC), Universidad Aut\'{o}noma de Madrid, Madrid, Spain}
\affiliation{Laboratorio de Transporte Cu\'{a}ntico, Unidad Asociada
UAM/ICMM-CSIC, Madrid, Spain}

\date{\today}

\begin{abstract}
Superconductor-to-metal transitions (SMTs) are key probes of mesoscopic superconductivity, but their interpretation can be complicated by device geometry and measurement conditions. Here, we study epitaxial InAs–Al nanowires and show that metallic contacts induce an inverse proximity effect (IPE), creating weak spots in the superconductor that strongly suppress the critical current and give rise to an anomalous metallic phase. Using transport measurements supported by Usadel theory, we demonstrate that this phase originates from the contact-induced weakening of superconductivity together with Joule heating, rather than intrinsic material properties. Our findings reveal an overlooked \textit{observer effect} in 
mesoscopic superconductors and provide essential guidance for interpreting SMTs and for designing  devices based on these systems.

\end{abstract}

\maketitle


The superconductor-to-metal transition (SMT) is widely employed 
as a tool to characterize the origin and properties of superconductors. While conventional bulk systems are well explained by BCS theory, interesting complications arise in mesoscopic samples. 
Here, device geometry and reduced dimensionality greatly impact both superconducting properties and the nature of SMTs~\cite{shah2007microscopic, sacépé_feigel’man_klapwijk_2020}, frequently resulting in complex regions of finite resistance dubbed \textit{anomalous metallic phases}~\cite{Kapitulnik2019Jan, wang_liu_ji_wang_2023}. In quasi-1D systems such resistance can originate from dynamical phase-slips~\cite{DelMaestro2009Mar}, while in quasi-2D it can arise from phase-slip lines formed by vortices~\cite{dmitriev_zolochevskii_2006, Zolochevskii2014Oct}. 
Other possible origins of such anomalous phases include granular or glassy superconductors, where inhomogeneities yield pockets of normal resistance or serve to suppress superconducting correlations altogether~\cite{spivak_a.zyuzin_hruska_2001, spivak_oreto_kivelson_2008, Kapitulnik2019Jan}. Recent experiments on 2D Josephson junction arrays have also demonstrated the phenomenon in artificial materials~\cite{bøttcher_nichele_kjaergaard_suominen_shabani_palmstrøm_marcus_2018, Sasmal_2025}. Although complex, these occurrences can shed light on the dynamics of quantum phase transitions, with practical implications for the development of theory and the interpretation of more unconventional superconductivity.

In this work, we study the superconducting properties of 
epitaxial InAs-Al nanowires by means of transport measurements, revealing different types of SMTs. 
Such hybrid superconductor-semiconductor systems \cite{Chang2015,Krogstrup2015} are receiving increasing interest for the development of novel quantum devices, e.g., gate-tunable hybrid superconducting qubits, such as gatemons \cite{Larsen2015,Danilenko2023}, hybrid fluxoniums \cite{pita2020gate}, and Andreev qubits \cite{Hays2021,PitaVidal2022,Cheung2024}, and in the search for topological superconductivity through both 1D \cite{lutchyn2010majorana, oreg2010helical, prada_andreev_2020} and quantum dot based approaches \cite{leijnse2012parity, sau2012realizing, Fulga_2013_QD, Dvir2023Feb}. Here, our main finding is a significant \textit{observer effect} in measured SMTs of 
superconducting InAs-Al nanowires, meaning that the method of probing significantly alters the system´s 
properties. More specifically, we 
demonstrate that the metallic leads used for electrically contacting the nanowires in a device create weak spots in the superconductor through the inverse proximity effect (IPE). This is observed here through a marked reduction of the critical current compared to the theoretical expectation due to pair-breaking \cite{Romijn_IcAl_1982, Anthore_DOS_SCwire_2003}. In addition, we show that these weak spots result in an anomalous metallic phase for certain ranges of electrical current, temperature, and magnetic field. Using the Usadel theory for diffusive superconductors~\cite{usadel_1970} in conjunction with IPE and Joule heating, we provide an explanation for the different phases of the nanowire, which shows consistency with the measured critical current values of the SMT.

Despite the intense activity involving hybrid superconductor-semiconductor nanowires in the past decade, such an \textit{observer effect} has remained largely 
overlooked. This includes a similar study of SMTs in InAs-Al wires in which an anomalous metallic phase 
was also measured and attributed to an intrinsic effect of the material \cite{vaitiekenas2020anomalous}. In contrast, our findings indicate that the anomalous metallic phase arises from the IPE, and in particular to a state in which regions underneath the metallic contacts become normal while the rest of the
wire remains superconducting. This mechanism is also able to explain the 
unusually low critical currents measured in similar nanowires by previous work \cite{vaitiekenas2020anomalous, vekris2021asymmetric, elalaily2023signatures}, 
in which the IPE from the leads is not accounted for. The purpose of this paper is to bring attention to this effect, underscoring ways to identify and analyze SMTs in inversely proximitized devices and providing guidelines for the design of quantum devices based on hybrid nanowires. 

\begin{figure}[h]
\includegraphics[width=1.0\linewidth]{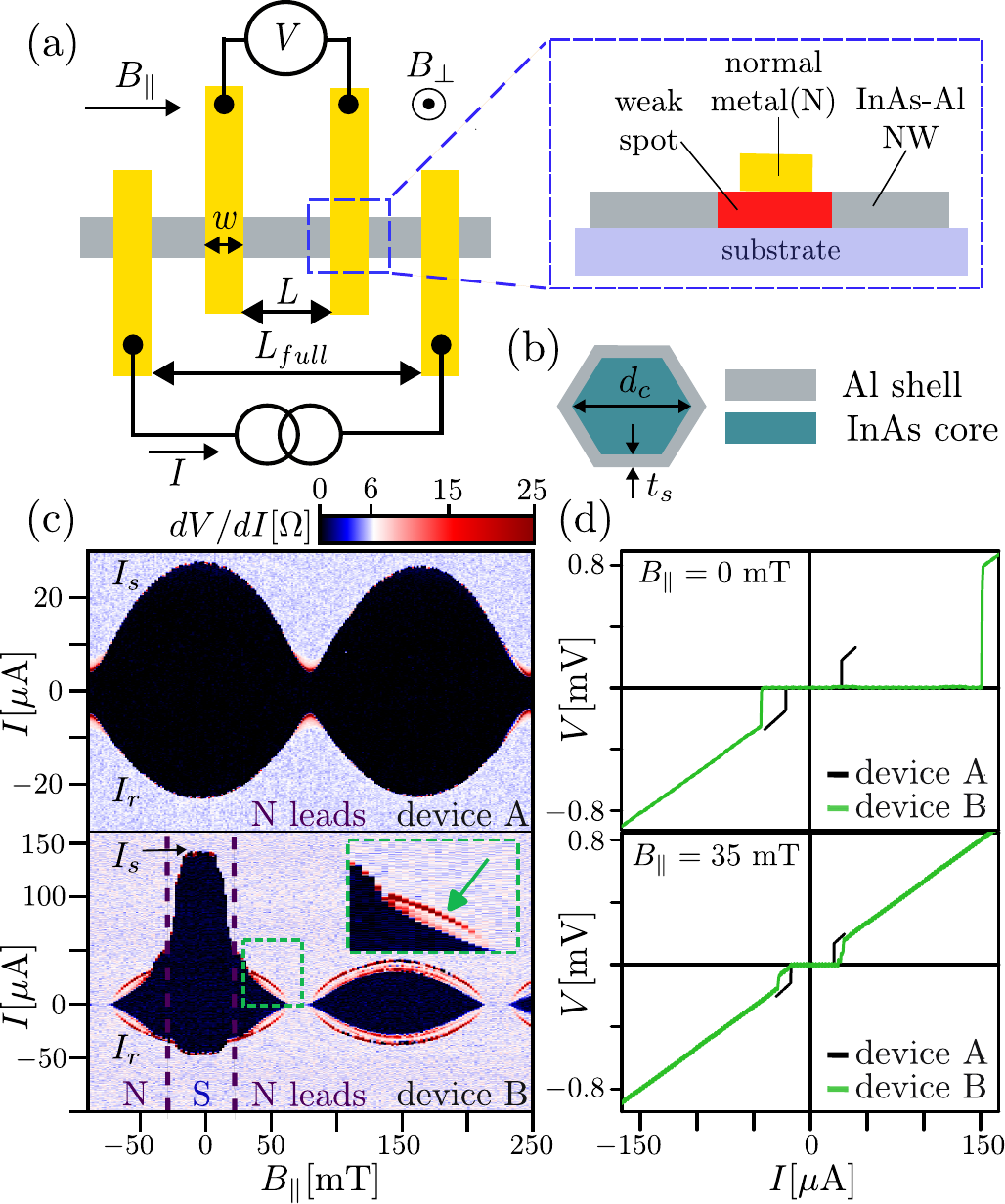}
\caption{\label{fig:1}
\textbf{Device geometry. (a)} Sketch of the four-terminal geometry with inner (outer) electrical contacts separated by a distance $L$ ($L_{full}$). The width of the electrical contacts is shown as $w$. The voltage drop, $V$, and the differential resistance, $dV/dI$ are measured as a response to a DC bias current, $I$, mixed with a small AC excitation. The inset schematically 
depicts a superconducting weak-spot induced in the superconducting 
wire by inverse proximity to a normal lead. \textbf{(b)} Schematic of the cross-section of a full-shell InAs-Al nanowire, with inner diameter, $d_c$, and Al shell thickness, $t_s$.
\textbf{(c)} Differential
resistance as a function of bias current and magnetic field applied parallel to nanowire axis for devices contacted by normal (device A) and superconducting leads (device B). 
The switching current, $I_s$, and retrapping current, $I_r$, oscillate due to the Little-Parks effect.
The vertical dashed lines in the lower panel mark the approximate critical field of the fabricated superconducting contacts, and the inset highlights a region of anomalous resistance.
\textbf{(d)} $V(I)$ traces taken at $B_\parallel = 0$~mT and $B_\parallel=35$~mT, showcasing the drastic reduction of $I_s$ and emergence of anomalous resistance in device B once the electrical contacts turn normal. 
}
\end{figure}

To study the impact of the metallic contacts on the SMTs of InAs-Al nanowires, we investigate devices with a four-terminal geometry, as sketched in Fig.~\ref{fig:1}(a). Measurements are taken by biasing a DC electrical current, $I$, mixed with a small low-frequency AC excitation between the two outermost contacts, and detecting the voltage drop, $V$, and the differential resistance, $dV/dI$, between the two inner leads (see Methods for more details). In this way, our experiment is sensitive to the properties of the nanowire segment of length $L$ located in between the voltage probes, including any possible IPE induced by them. To address this latter point, we explore devices with two different types of contacts: normal (Cr/Au, devices A and C) or superconducting (Ti/Al, device B).
Note that our InAs wires are fully covered by an Al shell (see schematic in Fig.~\ref{fig:1}(b)), which runs continuously throughout the length, $L_{full}$, of the device. For this reason, and owing to the much higher electron density in Al, we assume that $I$ flows entirely through the shell.

In Fig.~\ref{fig:1}(c), we show $dV/dI(I)$ measurements for devices A and B taken as a function of the magnetic field applied parallel to the nanowire axis, $B_{\parallel}$, and with $I$ being swept from negative to positive values. 
In these plots, we identify distinct values for the switching, $I_s$, and retrapping, $I_r$, currents which, respectively, denote superconductor-to-normal (S-N) and normal-to-superconductor (N-S) transitions. Note that such transitions manifest as peaks in differential resistance, reflecting the abrupt changes in $V$. The marked periodicities of $I_s(B_{\parallel})$ and $I_r(B_{\parallel})$ stem from the Little-Parks effect~\cite{Little1962Jul, vaitiekenas2020anomalous}, whereby an applied magnetic flux modulates the superconducting transition temperature of the Al shell due to fluxoid quantization. Strikingly, the switching currents for both devices differ substantially at zero field, i.e., $I_{s}^A(0) \approx 27$~$\mu$A and $I_{s}^B(0) \approx 150$~$\mu$A. At $B_{\parallel}\approx 25$~mT, corresponding to the critical field of the evaporated superconducting contacts, the switching current of device B drops sharply and, as shown in Figs.~\ref{fig:1}(c) and (d), reaches levels comparable to those of device A for higher fields (e.g., $I_s(B_{\parallel} = 35$~mT) $\sim 20~\mu$A for both devices). 
To shed light on these observations, it is instructive to consider the critical current of a diffusive 1D superconductor of length $L$ at zero magnetic field \cite{Romijn_IcAl_1982, Anthore_DOS_SCwire_2003},
\begin{equation}
I_{C}^{bulk} \approx 0.73\frac{\Delta_{BCS} L}{eR_N\xi_0},
\end{equation}
where $\Delta_{BCS}$ is the BCS superconducting gap, $\xi_0$ is the superconducting coherence length, and $R_N$ is the normal state resistance. By using parameters estimated for the Al shell in our wires (see Methods for details), we find $I^{bulk}_{C}\approx 150$~$\mu$A for both devices. 
We thus observe that $I_{s}(0)\approx I^{bulk}_{C}$ when the contacts are superconducting (device B), signaling that in this case the nanowire turns normal due to pair-breaking. In contrast, when the leads are normal as in device A or in device B for $|B_\parallel|>25$~mT, the switching current is strongly suppressed, which we attribute to the inverse proximity effect. More concretely, we relate the observed reduced switching currents to the critical current, $I_{C}$, of a superconducting weak spot formed under the normal contact due to IPE. Still, several observations
up to this point remain unclear, namely: (1) why does $I = I_{C}$ trigger a full S-N transition in device A (as evidenced by $dV/dI$ going to $R_N$)? (2) What is the origin of the hysteric behavior of the devices (i.e., $I_s\neq I_r$) and why is it sometimes not present (e.g., in  device B for $B_\parallel>25$~mT)? (3) What is the anomalous resistance state characterized by $0<dV/dI<R_N$, observed in device B for $B_\parallel>25$~mT when $I$ approaches $I_s$ (see inset in Fig.~\ref{fig:1}(c))?

To address the above questions, we
consider in more detail the different
transport regimes observed in our
superconducting nanowires. In Fig.~\ref{fig:2}(a), we show $dV/dI(I)$ measurements of device A now taken as a function of the magnetic field perpendicular to the wire, $B_{\perp}$. Instead of displaying oscillations,
$I_s$ and $I_r$ decrease rather monotonically with increasing field. Importantly, the characteristics of the nanowire SMT change at $B_{\perp}\approx40$~mT. For
lower fields than that, the device is hysteretic and its resistance changes abruptly from (to) zero to (from) $R_N$ at the S-N (N-S) transition. For $B_{\perp} \gtrapprox 40$~mT, the $dV/dI$ curves become symmetric with respect to $I$ and a region of anomalous resistance develops in between the zero- and normal-resistance states. To explain this behavior, we assume that the 
wire can be in one of three distinct 
phases, which we refer to as $\mathcal{N}$, $\mathcal{N}_S$, and $\mathcal{S}$. In $\mathcal{S}$, the entire wire is superconducting, including the weak spot formed by the IPE, and the device resistance, $R$, is zero. In $\mathcal{N}_S$, the weak spot has turned metallic while the remaining wire is still superconducting; this results in $0 < R < R_N$. Finally, in $\mathcal{N}$, the wire is fully metallic with $R=R_N$. In Fig.~\ref{fig:2}(b), a schematic of the different regions is shown in comparison to the experimental data. To gain a better understanding of the distinct phases of the superconducting wire, we assume that their stability is governed by two thresholds,
\begin{align}
I_{C}(B) &= I_{C}(0)\left(\frac{T_C(B)}{T_C(0)}\right)^{\gamma_{C}}, \label{eq:ICpower} \\
I_{T}(B) &= I_{T}(0)\left(\frac{T_C(B)}{T_C(0)}\right)^{\gamma_{T}}, \label{eq:ITpower}
\end{align}
which depend on powers of the superconducting critical temperature, $T_C(B)$ (see Supplemental Material for details).
Here, $I_C(B)$ describes the critical current of the weak spot, indicating when the $\mathcal{S}\rightarrow\mathcal{N}_S$ transitions occurs. $I_T(B)$ on the other hand, describes the $\mathcal{N}\leftrightarrow\mathcal{N_S}$ transition by balance of Joule heating, $P_h=IV$, with Wiedemann-Franz cooling~\cite{Ibabe2023May,Ibabe2024Jun}. It is thus the current at which the temperature of the central part of the wire, at a distance $L/2$ from the leads, cools below a superconductivity nucleation temperature, $T_N$, which we consider close to $T_C(B)$. Sketches of the two threshold mechanisms are shown in Fig.~\ref{fig:2} (c, d). Using Abrikosov-Gorkov theory (see Methods), we fit $I_s(B_{\perp})$ and $I_r(B_{\perp})$ in Fig.~\ref{fig:2} (a) to eqs.~(\ref{eq:ICpower},\ref{eq:ITpower}), obtaining good fits for $\gamma_C = 5/2$ and $\gamma_T = 1$ (yellow and cyan dashed lines). These values will be discussed in more detail later. 

\begin{figure}[t]
\includegraphics[width=1.0\linewidth]{Fig2_Gorm.pdf}
\caption{\label{fig:2}
\textbf{Regimes of inverse proximity effect in a superconducting wire. (a)} Differential 
resistance as a function of bias current and magnetic field applied perpendicular to nanowire axis. The yellow and cyan dashed lines indicate fits to eq.~(\ref{eq:ICpower}) and eq.~(\ref{eq:ITpower}) respectively, with $\gamma_C = 5/2$, $\gamma_T=1$. \textbf{(b)} Schematics of the the distinct nanowire phases, $\mathcal{N}$, $\mathcal{N}_S$, and $\mathcal{S}$, in the $I-B_{\perp}$ parameter space. A weak-to-strong IPE crossover occurs because $\gamma_C > \gamma_T$ and $I_C(0)>I_T(0)$. \textbf{(c)} Illustration of the mechanism setting $I_C$, where current is limited by the reduced gap under the metallic contact. \textbf{(d)} Depiction of the thermal balance 
between Joule heating and Wiedemann-Franz cooling governing $I_T$. 
}
\end{figure}

From the above, two distinct transport regimes can be defined based on whether $I_C(B)$ is greater or lower than $I_T(B)$. For $I_C(B) > I_T(B)$, the device is hysteretic and the only stable phases of the wire are $\mathcal{S}$ and $\mathcal{N}$. Indeed, when $I$ is swept 
up and reaches $I_C(B)$, the weak spot turns normal but $\mathcal{N}_S$ is unstable due to runaway heating \cite{Shah_thermalrunaway, Li_switchingAl_2008}. In this case, $P_h$ exceeds the cooling power of the device, and thus Joule heating quickly expands the normal region, increasing $R$ and $P_h=IV$ further until $\mathcal{N}$ is reached. By contrast, when the wire is in the $\mathcal{N}$ phase and $I$ is swept down, 
$\mathcal{S}$ becomes stable for $I = I_T(B)$ as superconducting nucleation leads to runaway cooling, shrinking the normal region and decreasing $P_h$ until $\mathcal{S}$ is stabilized. The behavior of device A for $B_{\perp} \lessapprox 40$~mT is captured by this first regime, which we dub the weak IPE regime. Conversely, for $I_T(B) > I_C(B)$, the device is non-hysteretic and the phase displaying anomalous resistance, $\mathcal{N}_S$, becomes allowed for certain regions of the parameter space ($I_T(B)>|I|>I_C(B)$). Here, the stability of $\mathcal{N}_S$ is ensured by thermal balance, which prevents runaway heating or cooling. This second regime, referred to as the strong IPE regime \footnote{Here, strong and weak refers to how much the IPE lowers $I_C$ when compared to $I_{C,0}$ and $I_T$.}, captures the behavior of device A for $B_{\perp} \gtrapprox 40$~mT. Note that the regime change observed in Fig.~\ref{fig:2} (a, b) originates from $\gamma_C>\gamma_T$,
and $I_C(0)>I_T(0)$, such that a cross-over occurs with increasing magnetic field.

We further assess our hypothesis of the distinct IPE regimes by means of Usadel models of 1D diffusive superconductors~\cite{usadel_1970, belzig_wilhelm_bruder_schön_zaikin_1999}, where we include the IPE in two finite-length segments of the wire~\cite{fominov_feigel’man_2001}. These calculations require two levels of numerical self-consistency~\cite{mooij_1982,vodolazov_2018} as both gap, $\Delta(x)$, and phase gradient, $\partial_x \phi(x)$, have to be obtained at each coordinate with an enforced DC current, $I$, as boundary condition (see Supplemental Material). Here, $I_C$ and $T_C$ are obtained as the threshold at which no stable superconducting solution exists. Again, we employ parameters estimated for the Al shell of our nanowires for the calculations. We identify two important scales in our models. First, the conductance between the normal contacts and the Al shell, $G_I$, sets the strength of the IPE and therefore plays a determining role on $I_C$. On the other hand, the separation between contacts, $L$, impacts $T_C$, as superconductivity in the wire becomes weaker for smaller contact separation. 

\begin{figure}[t]
\includegraphics[width=1.0\linewidth]{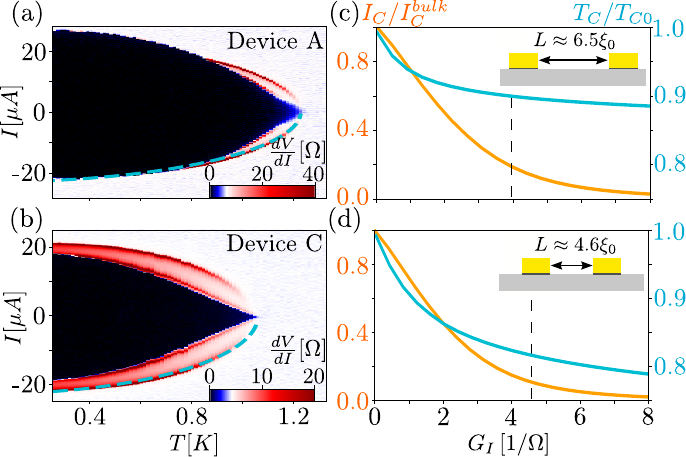}
\caption{\label{fig:3}
\textbf{Effect of the magnitude of the inverse proximity effect on $I_C$ and $T_C$. (a, b)} Differential 
resistance measurements of Devices A and C as a function of bias current and bath temperature, $T$. Cyan dashed lines displays $I_T(T) = I_T(0)\sqrt{T^2_C-T^2}$ used to extract $T_C^A$ and $T_C^C$. \textbf{(c, d)} Theoretically obtained estimates of $I_C/I_{C0}$ and $T_C/T_{C0}$ as a function of normal metal-superconductor interface conductance, $G_I$. Parameters are chosen to match Device A in (c) and Device C in (d).  
}
\end{figure}

To test the above behavior experimentally, we compare device A to a third device (device C), which also has normal contacts but smaller separation, $L$. In Fig.~\ref{fig:3} (a, b), we plot $dV/dI(I)$ of the devices as a function of the bath temperature, $T$. Although we estimate similar $I^{bulk}_{C}\approx 151$~$\mu$A for device C, the measured $I_{s}^C(0)$ is lower than $I_{s}^A(0)$. Also note that, unlike device A, the weak IPE regime is absent in device C. Aided by our Usadel models, we conclude that this behavior is a result of a stronger IPE in the latter device. Indeed, by fitting $G^A_{I} = 3.9$~S and $G^C_{I} = 4.6$~S for devices A and C, respectively, we capture the reduction of $I_C^{bulk}$ to the measured values of $I_s(0)$ as shown in Fig.~\ref{fig:3}(c, d). This also explains the lack of a weak IPE regime in device C, as it renders $I_T(0)>I_C(0)$.
Moreover, from the extracted values of $G_I$ and assuming an initial $T_{C0} = 1.4$~K of Al (see Methods), our models are able to predict the reduced superconducting transition temperatures of the devices, $T_{C}^A(0) \approx 1.26$~K and $T_{C}^C(0) \approx 1.14$~K, as seen by the cyan curves in Fig.~\ref{fig:3} (c, d).
These estimates are in good agreement with the measured $T_C^A\approx 1.23$~K and $T_C^C\approx 1.06$~K, thus supporting our analysis.

\begin{figure*}
\centering
\includegraphics[width=1.0\linewidth]{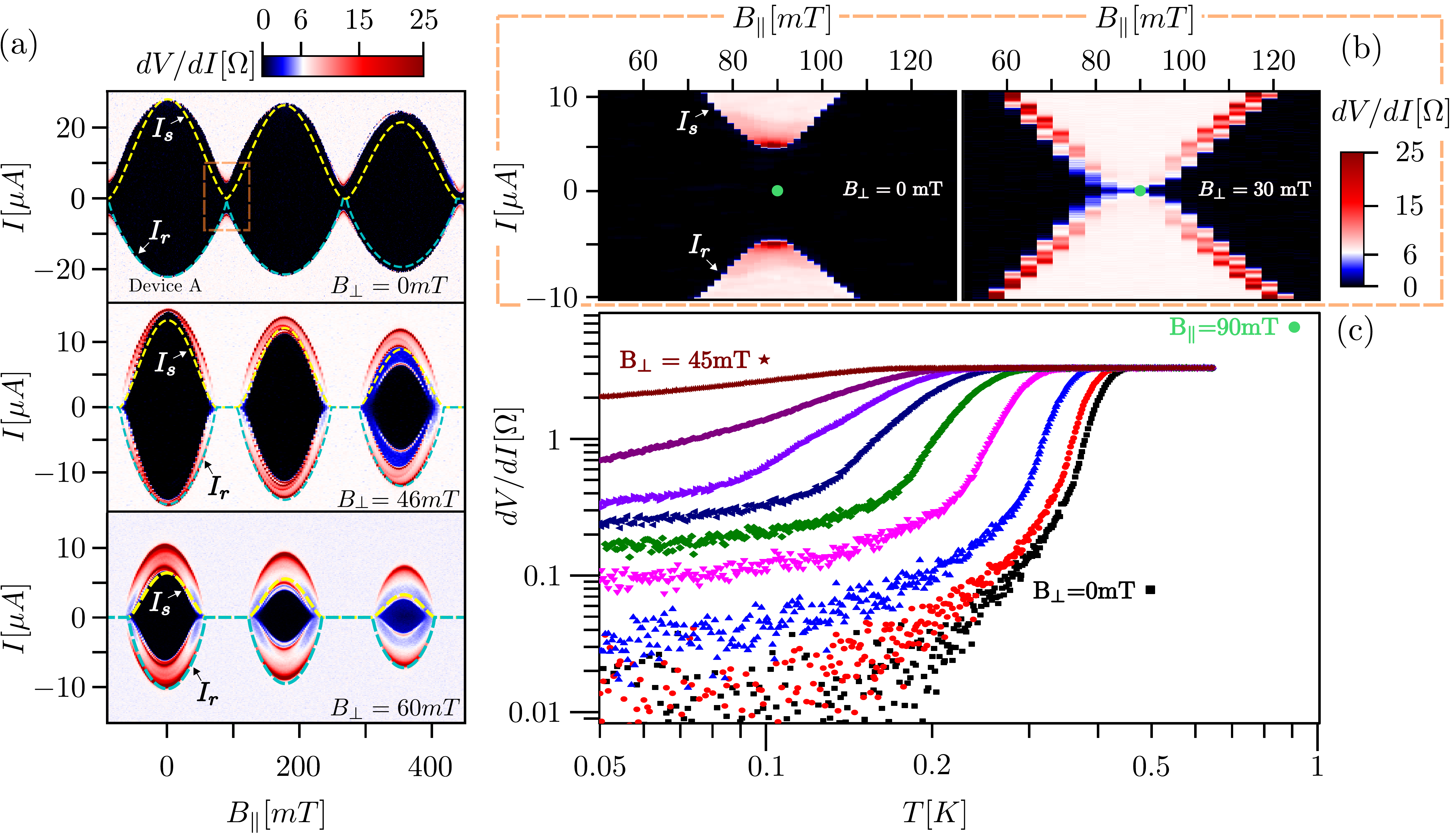}
\caption{\label{fig:4}
\textbf{Emergence of anomalous metallic phase with $B_{\perp}$.}  \textbf{(a)} $dV/dI(I, B_{\parallel})$ measurements of device A taken with $B_{\perp}$ = 0 (top panel), 46 (middle panel) and 60 mT (bottom panel). The yellow and cyan dashed lines correspond to $I_C(B)$ and $I_T(B)$ calculated using parameters obtained from the fit in Fig.~\ref{fig:2}(a) and $\gamma_C = 5/2$, $\gamma_T = 1$. \textbf{(b)} $dV/dI(I, B_{\parallel})$ near $\Phi = 0.5 \Phi_0$ with $B_{\perp}$ = 0 (left panel) and 30 mT (right panel). \textbf{(c)} Zero-bias $dV/dI$ taken at $\Phi = 0.5 \Phi_0$ ($B_{\parallel}$ = 90~mT, highlighted in panel (b) as a light green point) as a function of the bath temperature for various $B_{\perp}$ from 0 to 45 mT. 
}
\end{figure*}

By further including the pair-breaking from external magnetic fields
in our model, we are able to numerically calculate the $I_C(B)$ dependence of an inverse proximitized superconducting wire (see Supplemental Material for details). For the estimated $G_I$ values, we fit the numerically obtained $I_C(B)$ to a eq.~(\ref{eq:ICpower}) power-law, and find that theory supports $\gamma_C=5/2$ fitted to data (see yellow dashed lines in Figs.~\ref{fig:2} and ~\ref{fig:4}). Notably, this deviates from the expected $\gamma_C =3/2$ behavior of a non-proximitized superconductor. By its turn, we estimate $I_T(B)$ by considering that Joule heating is uniformly created in the metallic state of the wire and assuming that cooling occurs entirely through Wiedemann-Franz diffusion with the contacts anchored at $T$. We find,
\begin{equation}
P_T = I_TV_T = \frac{\pi^2}{3}\frac{k_B^2}{e^2}\frac{1}{R_h+4R_I}\left(T_N-T\right)^2,
\label{eq:PT}
\end{equation}
with $R_I = 1/G_I$, $V_T = R_hI_T$, $R_h =\frac{R_N}{2}\frac{w+d}{d}$, and $w$ being the contact width. Assuming that $T_N(B)\approx T_C(B)$ and $T_C \gg T$, we obtain a relation of the form of eq.~(\ref{eq:ITpower}) with $\gamma_T = 1$. This, again, agrees with fits to experiment in Figs.~\ref{fig:2} and \ref{fig:4}, and $I_T(T)$ in Fig.~\ref{fig:3} (a, b). In addition, from eq.~(\ref{eq:PT}),  we estimate $I_T^A(0) \approx 38~\mu$A and $I_T^C(0) \approx 41~\mu$A, which are approximately twice the measured values. We attribute this discrepancy to oversimplifications in the model related to the cooling mechanism and the nucleation of superconductivity in the metallic state.

Next, we explore in more detail the weak-spot transitions by tuning the IPE regime from weak to strong. This is accomplished by applying $B_{\perp}$ to modulate $I_C$, as shown in the Little-Parks plots of device A in Fig.~\ref{fig:4} (a). The top panel presents a measurement taken at $B_{\perp} = 0$, displaying up to the second Little-Parks lobe centered at $\approx350$~mT. Here, the overall behavior is consistent with the weak IPE regime, i.e., $I_s$ is a fraction of $I^{bulk}_{C}$, and the only stable phases are $\mathcal{S}$ and $\mathcal{N}$. As discussed in Fig.~\ref{fig:2}(a), the IPE regime in device A changes from weak to strong for $B_{\perp} \gtrapprox 40$~mT. Accordingly, in the middle and lower panels, we observe a growing stabilization of the $\mathcal{N}_S$ phase as $I_s$ is suppressed by $B_{\perp}$. The dashed yellow and cyan lines in all panels
correspond to $I_C(B)$ and $I_T(B)$ curves plotted using eqs.~(\ref{eq:ICpower}, \ref{eq:ITpower}) with the parameters obtained from the fit of Fig.~\ref{fig:2}(a). The agreement with the experimental data is remarkably good. We also note that in Fig.~\ref{fig:4} (a), it is possible to observe distinct intermediate resistance values in the $\mathcal{N}_S$ phase of lobes 1 and 2. We attribute this to the existence of two distinct weak spots related to the two contacts, which are likely asymmetric in $G_I$ and slightly misaligned to $B$ due to wire curvature, etc. 

To address the emergence of the $\mathcal{N}_S$ phase with $B_{\perp}$, we focus on measurements taken at half-flux quantum (Fig.~\ref{fig:4} (b, c)) and low currents. Near this condition, the Little-Parks oscillations become destructive with increasing $B_{\perp}$, meaning that normal regions emerge for $I=0$, owing to the overall weakening of the superconductor and of the weak spots. As of this weakening, metallic domains develop and expand around the weak spots, which consequently impact $I_C$ and lead to the emergence of intermediate resistance states. This manifests itself as a phase transition from a superconductor to an anomalous metal \cite{vaitiekenas2020anomalous}, whereby the resistance of the wire saturates at low temperatures to a $B_{\perp}$-dependent value intermediate between zero and $R_N$, even under conditions far from criticality ($T < T_C$ and $B_\perp < B_{\perp, C}$). 
This behavior is clearly observed in Fig.~\ref{fig:4} (c), where we plot zero-bias $dV/dI$ traces taken as a function of $T_B$ for different values of applied perpendicular field.

To conclude, our work reveals a strong effect of fabricated electrical contacts on the properties of superconducting nanowires. In particular, we have shown that the inverse superconducting proximity effect imparted by normal contacts results in the formation of weak spots that ultimately limit the critical current of a device. Our experiment, supported by Usadel models, therefore demonstrates that SMT measurements can unexpectedly reflect the properties of weak spots rather than those of the superconducting wire itself owing to a combination of IPE and Joule heating. We note that devices with high contact resistance (i.e., low $G_I$) may also experience premature switching even in the absence of IPE, due to the Joule effect and runaway heating. In addition, we uncover a mechanism for the anomalous metallic phase behavior reported for InAs-Al nanowires \cite{vaitiekenas2020anomalous} based on the stabilization of normal metallic regions near weak spots due to the IPE. Overall, this work has direct implications for superconducting devices incorporating normal electrical contacts and particularly for studies addressing the properties of mesoscopic superconductors  \cite{vaitiekenas2020anomalous, vekris2021asymmetric, elalaily2021gate, elalaily2023signatures}.
Our conclusions also have implications on device considerations for the realization 
of topological superconductors, as proximity effects from electrical contacts can result in variations in the superconducting gap along a nanowire.

\section*{Methods} \label{sec:Methods}
\textbf{Sample fabrication.} 
The superconducting wires studied here are full-shell InAs-Al nanowires with 
core diameters, $d_c$, of approximately 
105 nm for devices A and B, and 
125 nm for device C. The thickness of the Al shell, $t_S$ is estimated to be approximately 8 nm from fits to Little-Parks oscillations. The total nanowire diameter, $d_F$, is defined as $d_c + 2t_S$ (see Fig.~\ref{fig:1}(b)), while the total cross-sectional area of the wire, $A_F$, is given by $3\sqrt{3}d_F^2/8$.

For the fabrication, nanowires are deterministically transferred from the growth chip to Si/SiO$_{2}$(300$~$nm) substrates using a micromanipulator. 
Here, we have employed devices with a four-terminal geometry. Electrical contacts were fabricated using 
standard e-beam lithography (EBL) and lift-off techniques. Argon ion milling was employed prior to metalization to remove the native oxide from the Al shell. We have deposited two types of contacts by e-beam evaporation: (i) normal, Cr(2.5 nm)/Au(80 nm), for devices A and C, and (ii) superconducting, Ti(2.5 nm)/Al(240 nm), for device B. 
The width of the fabricated electrical leads, $w$, was fixed at 300$~$nm for all devices. The separation between innermost (outermost) contacts, i.e., $L$ ($L_{full}$) in Fig.~\ref{fig:1}, is equal to 650 nm (2.1~$\mu$m), 690 nm (2.3~$\mu$m) and 500 nm (2.1~$\mu$m) for devices A, B and C, respectively. 

In addition to the devices discussed in the manuscript, we have measured 2 devices with normal contacts and 3 devices with superconducting contacts. The reported phenomena have been overall reproduced on all devices.

\bigskip

\textbf{Measurements.} Device A (B) was measured in a dilution refrigerator operating at a base temperature of approximately 130~mK (100~mK). This system is equipped with a vector magnet, which was used to align the external magnetic field parallel or perpendicular to the nanowire axis. Device C, on the other hand, was measured using a $^3$He insert at a base temperature of $\approx 250$~mK. Current-bias transport measurements were carried out by means of standard low-frequency ($f$ = 107 Hz) lock-in techniques with excitations ranging from 60 to 200 nA.

\bigskip

\textbf{Estimation of parameters of the Al shell.} 
To analyze the Little-Parks oscillations in our experiment, we employ Abrikosov-Gorkov theory. 
The effect of a magnetic field on the nanowire is captured by the pair-breaking parameter, $\alpha=\alpha_\parallel+\alpha_\perp$, with the parallel component given by~\cite{shah2007microscopic},
\begin{equation}
    \alpha_{\parallel} = \frac{4\xi_0^2 k_B T_{C}(0)}{A_F} \left[\left( n - \frac{\Phi_{\parallel}}{\Phi_0} \right)^2 +  \frac{t_S^2}{d_F^2}\left( \frac{\Phi_{\parallel}^2}{\Phi_0^2} + \frac{n^2}{3}  \right)  \right],
    \label{eq:alpha_par}
\end{equation}
where $T_{C}(0)$ is the superconducting transition temperature at zero magnetic field,
$n$ is the fluxoid quantum number, $\Phi_{\parallel} = B_{\parallel}A_F$ is the applied flux parallel to the nanowire axis, and $\xi_0$ is the
superconducting coherence length.
The effect of an applied transverse magnetic field is given by,
\begin{equation}
    \alpha_{\perp}= \frac{4\xi_0^2 k_B T_{C}(0)\lambda}{A_F}\frac{\Phi_{\perp}^2}{\Phi_0^2}, 
    \label{eq:alpha_perp}
\end{equation} 
where $\Phi_{\perp} = B_{\perp}A_F$, and $\lambda$ is a free fitting parameter~\cite{vekris2021asymmetric}, which we find to be $\approx 1.3$. The parameter $\alpha$ has direct impact on the transition temperature of the superconductor, $T_C(\alpha)$, as described by the following Abrikosov-Gorkov equation:
\begin{equation}
    \ln\left( \frac{T_C(\alpha)}{T_{C}(0)} \right) = \Psi\left( \frac{1}{2} \right) - \Psi\left( \frac{1}{2} + \frac{\alpha}{2 \pi k_BT_C(\alpha)} \right),
    \label{eq: AG}
\end{equation}
where $\Psi$ is the digamma function. Using eqs.~(\ref{eq:ICpower}, \ref{eq:ITpower}) (with $\gamma_C = 5/2$ and $\gamma_T = 1$, respectively) together with eq.~(\ref{eq: AG}), we fit the measured $I_s(B)$ and $I_r(B)$. Considering $B_{\perp} = 0$, we use three fitting parameters for devices A and B: $\xi_0$, $A_F$ and $t_S$. For device C, we also include a fitting parameter for the angle between the magnetic field and nanowire, as this device was measured in a setup with a single-axis coil. For estimating $I_{C,0}$, we use $\xi_0$ from the Little-Parks fits and $R_N$ from the transport measurements for $I \gg I_s$.

The superconducting transition temperature of Al, $T_{C0} = 1.4$~K, used to estimate the reduced transition temperatures due to IPE, was obtained by Joule spectroscopy measurements \cite{Ibabe2023May} of devices employing similar nanowires.    

\bigskip

\textbf{Retrapping current estimate.} In this section we detail how eq.~(\ref{eq:PT}) is obtained. First, we assume that the wire is in the metallic state and ask at what power the middle part of the wire cools below $T_N$. From the center of one contact to the other is a length of $l+w$ in which $IV\approx2R_hI^2$ power is generated by Joule heating with $R_h=\frac{1}{2}\frac{d+w}{d}R_N$. In a symmetrical situation, half of this power runs to each contact via Wiedemann-Franz law. This is captured by the heating-cooling balance,
\begin{equation}
P(x)=-\frac{\pi^2}{3}\frac{k_B^2}{e^2}\sigma A_F T(x)\frac{\partial T}{\partial x}(x),
\end{equation}
where $P(x)=R_hI^2\frac{2x}{d+w}$ with $x=0$ defining the wire center, such that $P((d+w)/2)=R_h I^2$ underneath the contact center. Here, conductivity, $\sigma$, connects to normals resistance, $R_N=d/\sigma A_F$. Assuming the critical power is inputted via Joule heating, $P_T=R_hI_T^2$, the boundary conditions becomes $T(0)=T_N$ and $T((d+w)/2)=T_I$, and we integrate over $x$ to find,
\begin{equation}
P_T = \frac{\pi^2}{3}\frac{k_B^2}{e^2}\frac{1}{R_h}\left(T_N^2-T_I^2\right),
\end{equation}
where $T_I$ is the temperature underneath the contact center. Next, assuming that the contact receives symmetrical heat from the wire segment to the left and right of it, a total cooling power of $P_I=2P_T$ must cross the interface, $R_I$, given by,
\begin{equation}
P_I=2P_T = \frac{\pi^2}{6}\frac{k_B^2}{e^2}\frac{1}{R_I}\left(T_I^2-T^2\right).
\end{equation}
Combining these equations to remove the $T_I^2$ dependence yields eq.~(\ref{eq:PT}) of the main text.

\bibliography{bibtex}

\bigskip

\textbf{Author contributions}

A.I., G.M.O and I.C. fabricated the devices. G.M.O., I.C., and M.G. performed the measurements. G.M.O., G.O.S, I.C., M.G. and E.J.H.L. analyzed the experimental data. G.O.S., A.L.Y. and R.A. developed the theory. G.O.S. performed the theoretical calculations. T.K. and J.N. developed the nanowires. All authors discussed the results. G.O.S., G.M.O. and E.J.H.L, wrote the manuscript with input from all authors. 

\bigskip

\begin{acknowledgments}

We acknowledge funding from the EU through the European Research Council (ERC) Starting Grant agreement 716559 (TOPOQDot), the FET-Open contract 828948 (AndQC), the Danish National Research Foundation (DNRF 101), the SolidQ project of the Novo Nordisk Foundation, the Carlsberg Foundation, by the Spanish AEI through Grants No.~PID2020-117671GB-I00, PID2021-125343NB-I00, PID2023-150224NB-I00, TED2021-130292B-C41, and TED2021-130292B-C43, as well as through the ``Mar\'{\i}a de Maeztu'' Program for Units of Excellence in R\&D (CEX2018-000805-M) and the ``Severo Ochoa'' Centers of Excellence
program through Grant CEX2024-001445-S.

\end{acknowledgments}

\end{document}